# Interplay between the orbital quantization and Pauli effect in a charge-density-wave organic conductor[£]

Mark Kartsovnik,[a,*] Dieter Andres,[a] Pavel Grigoriev,[b,c] Werner Biberacher,[a] and Harald Müller[d]

[a]*Walther-Meißner-Institut, Bayerische Akademie der Wissenschaften, Walther-Meißner-Str. 8, D-85748 Garching, Germany*

[b]*L.D. Landau Institute for Theoretical Physics, Russian Academy of Sciences, 142432 Chernogolovka, Russian Federation*

[c]*High Magnetic Field Laboratory, MPI-FKF and CNRS, B.P. 166, F-38042 Grenoble, Cedex 09, France*

[d]*European Synchrotron Radiation Facility, B.P. 220, F-38043 Grenoble, France*

**Abstract**

The interlayer magnetoresistance of the low-dimensional organic metal $\alpha$-(BEDT-TTF)$_2$KHg(SCN)$_4$ under pressure shows features which are likely associated with theoretically predicted field-induced charge-density-wave (FICDW) transitions. At ambient pressure, a magnetic field strongly tilted towards the conducting layers induces a series of hysteretic anomalies. We attribute these anomalies to a novel kind of FICDW originating from a superposition of the orbital quantization of the nesting vector and Pauli effect on the charge-density wave. © 2001 Elsevier Science. All rights reserved

*Keywords:* Organic conductors; $\alpha$-(BEDT-TTF)$_2$KHg(SCN)$_4$; charge-density wave; field-induced charge-density waves; orbital quantization

*PACS:* 71.45.Lr; 72.15.Gd; 71.30.+h; 72.80.Le

## 1. Introduction

The layered organic metal $\alpha$-(BEDT-TTF)$_2$KHg(SCN)$_4$ has been of high interest over the last decade due to its low-temperature state showing an unusual behavior at high magnetic fields (see e.g. [1] for a review). At present there exist a number of strong experimental arguments] that this state is determined by a charge-density-wave (CDW) due to nesting of the quasi-one-dimensional (q1D) Fermi surface (FS). The critical temperature of the CDW transition, $T_c \approx$ 8.5 K, and, therefore, the relevant energy gap are much smaller than it is usually met in CDW materials. One of important consequences of this fact is a very strong influence of a magnetic field on electronic properties.

Generally, two different mechanisms of coupling of a magnetic field to a CDW should be considered. On the one hand, the Pauli response of spins of the interacting electrons leads to a gradual suppression of the CDW. The relevant "magnetic field – temperature" ($B$–$T$) phase diagram is largely similar to that of a clean superconductor or a spin-Peierls system [8]. In particular, it includes the high-field, low-temperature phase CDW$_x$ corresponding, respectively, to the Larkin-Ovchinnikov-Fulde-Ferrel or soliton lattice phases in the latter two cases. It is mainly the *Pauli effect* which determines the shape of the $B$-$T$ phase diagram of $\alpha$-(BEDT-TTF)$_2$KHg(SCN)$_4$ at ambient pressure, in magnetic field perpendicular to the highly conducting *ac*-plane [2-4].

While the Pauli effect is operative in any CDW system, provided the field is strong enough, the second, *orbital*

---

*effect* becomes important when the FS nesting is sufficiently imperfect. In layered conductors, the imperfectness of nesting can be introduced by a finite second-order interchain transfer integral $t'_\perp$ describing the next-nearest-chain transfer in the plane of conducting layers. If $t'_\perp$ is comparable to the characteristic CDW energy $t^* = \Delta_0/2$ ($\Delta_0$ is the CDW gap at $t'_\perp=0$), the zero-field critical temperature $T_c(0)$ becomes considerably lower than in the case of a perfectly nested FS. However, a magnetic field applied perpendicular to the layers restricts electron orbits to the chain direction, i.e. enhances their one-dimensionality, and restores the density-wave state. In particular, if $t'_\perp \geq t'^*$, the system is predicted [9,10] to undergo at low temperatures a series of first-order phase transitions to field-induced CDW (FICDW) subphases with quantized values of the $Q_x$ component of the nesting vector, in analogy with the known field-induced spin-density waves (FISDW) phenomenon. However, in the CDW case one should take into account a competition between the Pauli and orbital effects.

While the classical orbital effect has been shown to exist in the CDW compound α-(BEDT-TTF)$_2$KHg(SCN)$_4$ under pressure, only weak signs for possible FICDW transitions were observed [4]. In the next section, we present data on the magnetoresistance under quasi-hydrostatic pressure which further support the existence of FICDW in this material. In section 3, we argue that a new manifestation of the orbital quantization which originates from *simultaneous* effects of Pauli and orbital coupling of a high magnetic field to a CDW is observed in α-(BEDT-TTF)$_2$KHg(SCN)$_4$ in strongly tilted fields already at ambient pressure.

## 2. Magnetoresistance in perpendicular magnetic fields, under pressure

Figure 1 shows interlayer magnetoresistance of α-(BEDT-TTF)$_2$KHg(SCN)$_4$ in fields perpendicular to the conducting layers recorded at $T = 0.1$ K at different pressures. Strong Shubnikov-de Haas (SdH) oscillations with the fundamental frequency of 670 to 740 T (depending on pressure) originate from the second, q2D band which remains metallic in the CDW state. At low pressure these oscillations are superimposed on a very high smooth background typical of the CDW state of this compound. With increasing $p$ the CDW state is gradually suppressed which is, in particular, reflected in a rapid decrease of the background magnetoresistance. Further, when the pressure exceeds the critical value $p_c \cong 2.3$-$2.5$ kbar at which the zero-field CDW completely vanishes [4,11], new oscillatory features emerge in the $R(B)$ curves. They are most prominent at $p = 3.5\pm0.5$ kbar and fade away outside this pressure interval. It is important that this interval exactly matches the conditions at which the FICDW transitions are expected for this compound [4,9]: it corresponds to the antinesting parameter $t'_\perp$ exceeding but still close to the critical value $t'^*$. At $t'_\perp < t'^*$ the CDW state is already established at zero field. On the other hand, at $t'_\perp$ considerably larger than $t'^*$ the negative influence of the Pauli effect on the CDW overwhelms the orbital effect [9].

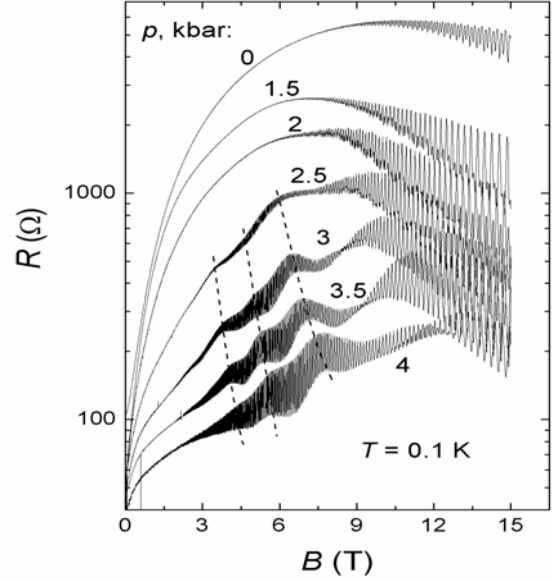

Fig. 1. Magnetoresistance of α-(BEDT-TTF)$_2$KHg(SCN)$_4$ in perpendicular fields at different pressures.

While the frequency of the observed features does not change significantly between 2.5 and 4 kbar, their positions shift notably to higher fields with increasing pressure (see dashed lines in Fig. 1), in line with their proposed FICDW nature. Indeed, with increasing $t'_\perp$ one needs a higher field to induce a CDW state via the orbital effect.

The magnitude of the features rapidly decreases with increasing temperature (see Fig. 2), vanishing above ~1.4 K. This agrees with Lebed's prediction that the FICDW transitions should emerge at considerably lower temperatures than FISDW [10]. Finally, as seen in Fig. 2, the positions of the features notably depends on temperature in a way similar to that observed in the FISDW case, in contrast to what one would normally expect from a usual SdH effect.

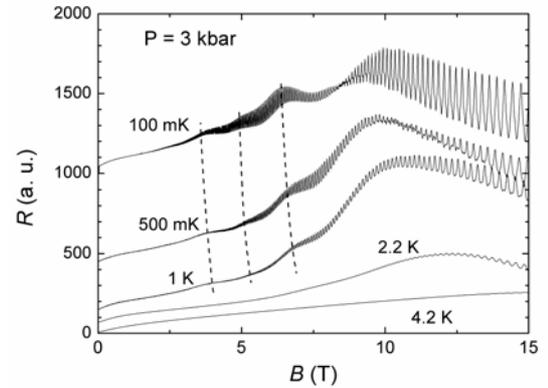

Fig. 2. Field sweeps of the magnetoresistance at different temperatures; $p = 3$ kbar.

## 3. Ambient-pressure FICDW transitions in strongly tilted fields

We now consider the influence of a high magnetic field strongly tilted towards the conducting plane of $\alpha$-(BEDT-TTF)$_2$KHg(SCN)$_4$ on the CDW state at ambient pressure. Figure 3 shows magnetoresistance (a) and magnetic torque (b) at different field orientations. The orientation is defined by a tilt angle $\theta$ between the field and the normal to the conducting $ac$-plane and azimuthal angle $\varphi$ between the field projection on the $ac$-plane and the $c$-axis (see inset in Fig. 3). All the curves display a complicated hysteretic structure consistent with previous reports [3,12].The hysteretic character of the anomalies suggests that they are associated with multiple first order phase transitions. Fig. 3 illustrates that the positions of the anomalies are independent of the azimuthal angle $\varphi$. On the other hand, they are known to strongly depend on the tilt angle $\theta$ [13]. Fig. 4 shows the positions of local maxima of the torque derivative $(\partial\tau/\partial B)_\theta$ versus angle $\theta$. Crosses correspond to the so-called kink field $B_k$ which is associated with the transition between the low-field state CDW$_0$ and high-field state CDW$_x$. At $\theta \leq 40°$ the kink field is constant, $B_k \approx 23$ T; at higher $\theta$ it starts moving to lower fields. The anomalies above $B_k$ emerge at $\theta \geq 65°$ and also rapidly shift down, approaching $B_k$ at $\theta \to 90°$. The increasingly high sensitivity of the structure to changes in the tilt angle $\theta$ near 90° and its independence of the azimuthal orientation $\varphi$ suggests that the orbital effect determined by the field component $B_z = B\cos\theta$ plays a crucial role. On the other hand, it is important that the anomalies occur at $B>B_k$, i.e. in the fields producing a very strong Pauli effect. Therefore it is very likely that the observed transitions originate from an interplay between the Pauli and orbital effects on the high field CDW state.

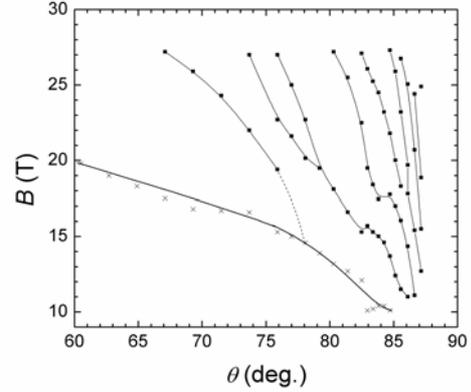

Fig. 4. $B$–$\theta$ phase diagram of $\alpha$-(BEDT-TTF)$_2$KHg(SCN)$_4$. Crosses and solid symbols are, respectively, the kink transition $B_k$ and the multiple transitions within the high-field state.

In order to qualitatively understand the origin of the new transitions, we consider the field dependence of the $Q_x$ component of the nesting vector in a CDW system with a moderately imperfect nesting ($t'_\perp < t'^*$). At zero field, $Q_{0x} \approx 2k_F$ corresponds to the optimal nesting; the entire FS is gapped. At a finite magnetic field the degeneracy between the CDW's with different spin orientations is lifted. Treating each spin subband independently, one can express the optimal nesting conditions as $Q_{\text{opt},x}(B) = Q_{0x} \pm 2\mu_B B/\hbar v_F$ where $\mu_B$ is Bohr magneton, $v_F$ is the Fermi velocity in the chain direction and the sign +(-) stands for the spins parallel (antiparallel) to the applied field. This splitting of the optimal nesting conditions is illustrated by dashed lines in Fig. 5. Nevertheless, both subbands remain fully gapped and the system as a whole maintains the constant nesting vector $Q_{0x}$ up to the critical field $B_k \sim \Delta(B=0)/2\mu_B$. Above $B_k$, $Q_{0x}$ is no more a good nesting vector as it leads to ungapped states in both subbands. As shown by Zanchi et al. [9], the CDW energy can be minimized in this case by introducing a field dependent term $q_x^{\text{Pauli}} = Q_x(B) - Q_{0x}$ which is schematically represented in Fig. 5 by the thin solid line asymptotically approaching the value $2\mu_B B/\hbar v_F$. This obviously improves the nesting conditions for one of the spin subbands (say, the spin-up subband) at the cost of an additional "unnesting" of the other (spin-down).

Now it is important to take into account that the spin-down subband becomes unnested at $B > B_k$ and therefore is subject to a strong orbital effect. The situation is analogous to that with a large antinesting term $t'_\perp \geq t'^*$. One can therefore expect that, like in the "conventional" FISDW or FICDW case, an orbital quantization condition be set on the system. However, unlike in the FISDW case, the quantized levels are counted from $(Q_{0x} - 2\mu_B B/\hbar v_F)$ rather than from $Q_{0x}$. The corresponding values $q_{xN} = -2\mu_B B/\hbar v_F + NG$ (where $G = ea_y B_z/\hbar c$) are shown by dotted lines in Fig. 5.

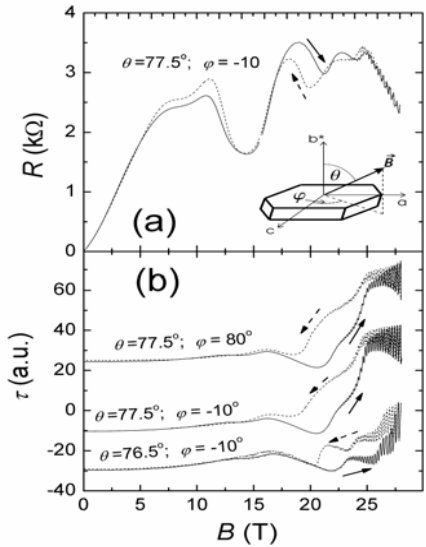

Fig. 3. Resistance (a) and magnetic torque (b) versus magnetic field strongly tilted towards the $ac$-plane. The angles $\theta$ and $\varphi$ define the field orientation as shown in the inset.

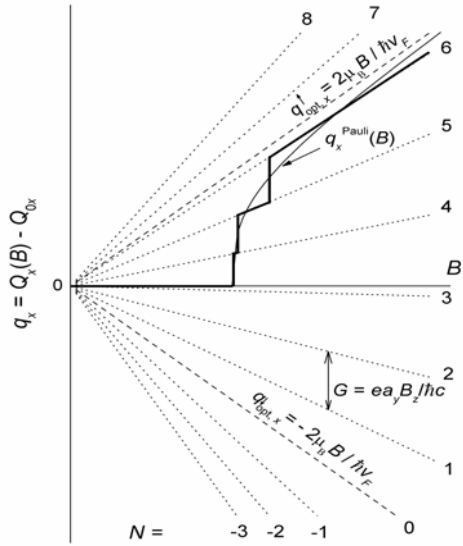

Fig. 5. Schematic illustration of the superposition of the Pauli effect and orbital quantization on the CDW nesting vector.

As a result, the most favorable values of the nesting vector above $B_k$ are determined by intersections of the continuous curve $q_x^{Pauli}$ with the straight lines $q_{xN}$, i.e. by the superposition of the Pauli and quantum orbital effects. Thus, with changing the field we obtain a series of first order transitions between CDW subphases characterized by different quantized values of the nesting vector as schematically shown by thick lines in Fig. 5. Note that the quantum number *N increases* with the field, in contrast to what is usually observed in known orbital quantization phenomena.

The multiple FICDW transitions can be observed when the distance $G$ between the quantized levels is smaller than $\mu_B B/\hbar v_F$. This condition is obviously not fulfilled for $\alpha$-(BEDT-TTF)$_2$KHg(SCN)$_4$ at the field perpendicular to the layers. With tilting the field, $G$ reduces, being determined by $B_z = B\cos\theta$, whereas the Pauli effect remains unchanged. This causes the transitions at low enough $\cos\theta$. With further increasing $\theta$, the transitions shift to lower fields, in agreement with the experiment.

Thus, the presented qualitative model seems to explain the physical origin of the multiple field-induced transitions in $\alpha$-(BEDT-TTF)$_2$KHg(SCN)$_4$ and their evolution with changing the field orientation. The real phase lines shown in Fig. 4 look somewhat more complicated than one would derive from this simple consideration. A more thorough theoretical analysis aimed to provide a quantitative description of the new phenomenon is in progress.

## 4. Acknowledgements

The experiments in fields above 15 T were performed in Grenoble High Magnetic Field Laboratory. The work was supported by HPP Programme of EU, contracts HPRI-1999-CT-00030 and HPRI-CT-1999-40013, INTAS grant 01-0791, and RFBR 03-02-16121.

**References**

[1] M. Kartsovnik, V. Laukhin, J. Phys. I France 6 (1996) 1753.
[2] M. Kartsovnik et al., Synth. Metals 86 (1997) 1933; N. Biskup et al., Solid State Commun. 107 (1998) 503.
[3] P. Christ et al., JETP Lett. 71 (2000) 303.
[4] D. Andres et al., Phys. Rev. B 64 (2001) 161104.
[5] N. Harrison, Phys. Rev. B 66 (2002) 121101.
[6] M. Basletic et al., Synth. Metals 120 (2001) 1021; B. Dora et al., Phys. Rev. B 66 (2002) 165116.
[7] P. Foury-Leylekian et al., Synth. Metals 137 (2003) 1271.
[8] A. Buzdin, V. Tugushev, Sov. Phys. JETP 58 (1984) 428.
[9] D. Zanchi, A. Bjelis, G. Montambaux, Phys. Rev. B 53 (1996) 1240.
[10] A. Lebed, JETP Lett. 78 (2003) 138.
[11] D. Andres et al., J. Phys. IV France 12 (2002) Pr9-87.
[12] P. Christ et al., Synth. Metals 70 (1996) 823.
[13] M. Kartsovnik et al., Synth. Metals 120 (2001) 687.